\documentclass[iop]{emulateapj}

\usepackage{color}

\slugcomment {}

\shorttitle{WISEA J114724.10$-$204021.3: A Planetary Mass Member of the TW Hya Association}
\shortauthors{Schneider et al.}

\begin{document}
\title{WISEA J114724.10$-$204021.3: A Free-Floating Planetary Mass Member of the TW Hya Association}

\author{Adam C. Schneider\altaffilmark{1,4}, James Windsor\altaffilmark{1}, Michael C. Cushing\altaffilmark{1}, J. Davy Kirkpatrick\altaffilmark{2}, \& Edward L. Wright\altaffilmark{3}}  

\altaffiltext{1}{Department of Physics and Astronomy, University of Toledo, 2801 W. Bancroft St., Toledo, OH 43606, USA; Adam.Schneider@Utoledo.edu}
\altaffiltext{2}{Infrared Processing and Analysis Center, MS 100-22, California Institute of Technology, Pasadena, CA 91125, USA}
\altaffiltext{3}{Department of Physics and Astronomy, UCLA, 430 Portola Plaza, Box 951547, Los Angeles, CA 90095-1547, USA}
\altaffiltext{4}{Visiting Astronomer at the Infrared Telescope Facility, which is operated by the University of Hawaii under contract NNH14CK55B with the National Aeronautics and Space Administration.}

\begin{abstract}
We present WISEA J114724.10$-$204021.3, a young, low-mass, high probability member of the TW Hya association.  WISEA J114724.10$-$204021.3 was discovered based on its red AllWISE color (W1$-$W2 = 0.63 mag) and extremely red 2MASS $J-K_{\rm S}$ color ($>$ 2.64 mag), the latter of which is confirmed with near-infrared photometry from the VISTA Hemisphere Survey ($J-K_{\rm S}$ = 2.57$\pm$0.03). Follow-up near-infrared spectroscopy shows a spectral type of L7 $\pm$ 1 as well as several spectroscopic indicators of youth. These include a peaked $H$-band shape and a steeper $K$-band slope, traits typically attributed to low surface gravity.  The sky position, proper motion, and distance estimates of WISEA J114724.10$-$204021.3 are all consistent with membership in the $\sim$10 Myr old TW Hya association.  Using the age of the TW Hya association and evolutionary models, we estimate the mass of WISEA J114724.10$-$204021.3 to be 5$-$13 $M_{\rm Jup}$, making it one of the youngest and lowest mass free-floating objects yet discovered in the Solar neighborhood.        

\end{abstract}


\section{Introduction}
Young, late-type L dwarfs (brown dwarfs with ages $<$ 100 Myr and $T_{\rm eff}$ $\lesssim$ 1600 K) show striking spectroscopic similarities to young, directly-imaged exoplanets (e.g., \citealt{gizis12}, \citealt{fah13}, \citealt{liu13}). Unlike exoplanets, however, free-floating brown dwarfs are much simpler to observe because they lack a nearby, bright host sun. As such, young, low-mass brown dwarfs are ideal laboratories for investigating the physical conditions likely to be present in giant exoplanets, thus offering critical checks of theory. While young brown dwarfs belonging to nearby (d $<$100 pc) young associations are beginning to be found in greater numbers (e.g., \citealt{gagne15}), very few young, late-type L ($>$L5) dwarfs are currently known ($\lesssim$10; see Figure 4 of \citealt{gagne15}).  However, it is these young, late-type L dwarfs that are crucial as comparison objects for directly imaged young brown dwarf and planetary companions (e.g., \citealt{bowl14}, \citealt{gauza15}, \citealt{hink15}, \citealt{bonn15}).

If a substellar object can be connected to a young, nearby moving group, then its age can be firmly established, thereby providing a vital anchor point for low-mass evolutionary models.  The young (10 $\pm$ 3 Myr -- \citealt{bell15}) TW Hya association (TWA) is one of the nearest regions of recent star formation.  Its proximity ($\sim$50 pc) and young age make it an excellent testbed for studying early phases of stellar and substellar evolution.  Here we report the discovery of WISEA J114724.10$-$204021.3, a new L7 member of the TW Hya association.       

\section{Identification of WISEA J114724.10$-$204021.3}

As noted in \cite{schneid14}, young, late-type L dwarfs occupy a unique region of 2 Micron All-Sky Survey (2MASS; \citealt{skrut06}) and Wide-field Infrared Survey Explorer ({\it WISE}; \citealt{wright10}) color space compared to field L dwarfs because of their extremely red near-infrared colors (see their Figure 5).  WISEA J114724.10$-$204021.3 (hereafter WISEA 1147$-$2040) was found as part of a larger program focused on finding young, late-type L dwarfs based on their 2MASS and AllWISE\footnote{http://http://wise2.ipac.caltech.edu/docs/release/allwise/expsup/} colors. Briefly, candidate young L dwarfs were chosen by requiring that they have a 2MASS $J - K_{\rm S}$ color between 2.0 and 3.5 mag and an AllWISE W1 (3.6 $\mu$m) $-$ W2 (4.5 $\mu$m) color between 0.3 and 0.9 mag.  Candidates were also required to have their $K_{\rm S}$, W1, and W2 passband uncertainties be greater than zero (i.e., not upper limits).  Note that we did not require objects to be well detected in the $J$-band, anticipating the existence of objects so red in $J - K_{\rm S}$ color that they may be detected at the $K_{\rm S}$ band, but not at $J$ (as is the case for WISEA 1147$-$2040).  Lastly, we required that the separation between the 2MASS and AllWISE source positions of candidates be greater than 1$\arcsec$, thereby ensuring each candidate shows appreciable proper motion between the 2MASS and AllWISE epochs ($\gtrsim$100 mas yr$^{-1}$, considering the $\sim$10 year time baseline between 2MASS and AllWISE).  We then scrutinized each candidate individually by inspecting available optical (DSS and SDSS), near-infrared (2MASS), and mid-infrared (AllWISE) images to ensure each candidate is a point source (i.e., not extended or blended) with noticeable proper motion.  Of the $\sim$50 returned candidate red L dwarfs, WISEA 1147$-$2040 was picked out early as a strong young L dwarf candidate because of its uncommonly red 2MASS $J - K_{\rm S}$ color ($>$2.64 mag) and therefore worthy of follow-up spectroscopic observations.  The basic properties and photometry of WISEA 1147$-$2040 are given in Table 1.

\begin{deluxetable}{lccc}
\tablecaption{WISEA J114724.10$-$204021.3 Properties}
\tablehead{
\colhead{Parameter} & \colhead{Value} & \colhead{Ref.}}
\startdata
\cutinhead{Identifiers}
AllWISE & J114724.10-204021.3 & 1\\
2MASS & 11472421$-$2040204 & 2\\
\cutinhead{Observed Properties}
$\alpha$ (J2000) & 11:47:24.10 & 1\\ 
$\delta$ (J2000) & $-$20:40:21.3 & 1\\ 
$\mu$$_{\alpha}$ & $-$122.1$\pm$12.0 mas yr$^{-1}$ & 4\\
$\mu$$_{\delta}$  & $-$74.5$\pm$11.3 mas yr$^{-1}$ & 4\\
$J$ (2MASS) & $>$17.511 mag & 2\\
$H$ (2MASS) & 15.764 $\pm$ 0.112 mag & 2\\
$K_{\rm S}$ (2MASS) & 14.872 $\pm$ 0.106 mag & 2\\
$Y$ (VHS) & 19.160 $\pm$ 0.067 mag & 3\\
$J$ (VHS) & 17.445 $\pm$ 0.029 mag & 3\\
$K_{\rm S}$ (VHS) & 14.872 $\pm$ 0.011 mag & 3\\
W1 & 13.718 $\pm$ 0.026  mag & 1\\
W2  & 13.090 $\pm$ 0.030 mag & 1\\
W3  & $>$12.155 mag & 1\\
W4  & $>$8.913 mag & 1\\
$J-K_{\rm S}$ (2MASS) & $>$2.64 mag & 2\\
$J-K_{\rm S}$ (VHS) & 2.57 $\pm$ 0.03 mag & 3\\
$J-K_{\rm S}$ (synthetic)  & 2.61 $\pm$ 0.03 mag & 4\\
$W1-W2$  & 0.63 $\pm$ 0.04 mag & 1\\
\cutinhead{Inferred Properties}
Spectral Type (NIR) & L7 $\pm$ 1 (red) & 4\\
Photometric Distance & 31.2 $\pm$ 1.5 pc & 4\\
Kinematic Distance & 32$-$33 pc & 4\\
Predicted Distance &  31.3 $\pm$ 3.8 pc & 4,5\\
$T_{\rm eff}$\tablenotemark{a}  & 1500 $\pm$ 100 K  & 4\\
log g\tablenotemark{a}  & 4.0 $\pm$ 0.5 & 4\\
Mass\tablenotemark{a}  & 6$-$13 $M_{\rm Jup}$ & 4\\
$T_{\rm eff}$\tablenotemark{b} & 1100$-$1200 K & 4\\
Mass\tablenotemark{b}  & 5$-$6 $M_{\rm Jup}$ & 4\\
Age  & 10 $\pm$ 3 Myr & 6
\enddata
\tablenotetext{a}{From model fitting (Section 4.2).}
\tablenotetext{b}{From the estimated bolometric luminosity (Section 5).}
\tablerefs{ (1) AllWISE; (2) 2MASS; (3) VHS; (4) This work; (5) BANYAN II (\citealt{malo13}, \citealt{gagne14}); (6) \cite{bell15}}

\end{deluxetable}

\section{Observations}

\subsection{IRTF/SpeX}
WISEA 1147$-$2040 was observed with the upgraded SpeX spectrograph \citep{ray03} on the night of 2016 Feb 12 UT at the NASA Infrared Telescope Facility (IRTF) on Mauna Kea.  The observations were made in prism mode with a 0$\farcs$5 wide slit, which achieves a resolving power ($\lambda$/$\Delta\lambda$) of $\sim$150 over the range 0.8 - 2.5 $\mu$m.  We oriented the 15$\arcsec$ long slit along the parallactic angle and obtained a series of 18 120s exposures at two different nod positions along the slit for a total exposure time of 2160s.  The A0V star HD 101122 was observed at a similar airmass for telluric correction purposes.  The spectrum was reduced using the SpeXtool reduction package (\citealt{cush04}; \citealt{vacca03}).  The extracted spectrum was flux calibrated using the VHS $K_{\rm S}$ photometry.    

\subsection{VISTA/VHS}
WISEA 1147$-$2040 was observed by the Visible and Infrared Survey Telescope for Astronomy (VISTA; \citealt{emer04}) in the $Y$, $J$, and $K_{\rm S}$ photometric bands (Figure 1) as part of the VISTA Hemisphere Survey (VHS; PI McMahon, Cambridge, UK).  WISEA 1147$-$2040 was well detected in all three bands and  the photometry from VHS is listed in Table 1.  These observations confirm the very red $J-K_{\rm S}$ color of WISEA 1147$-$2040 ($J-K_{\rm S}$ = 2.57 $\pm$ 0.03 mag).  

\begin{figure}
\plotone{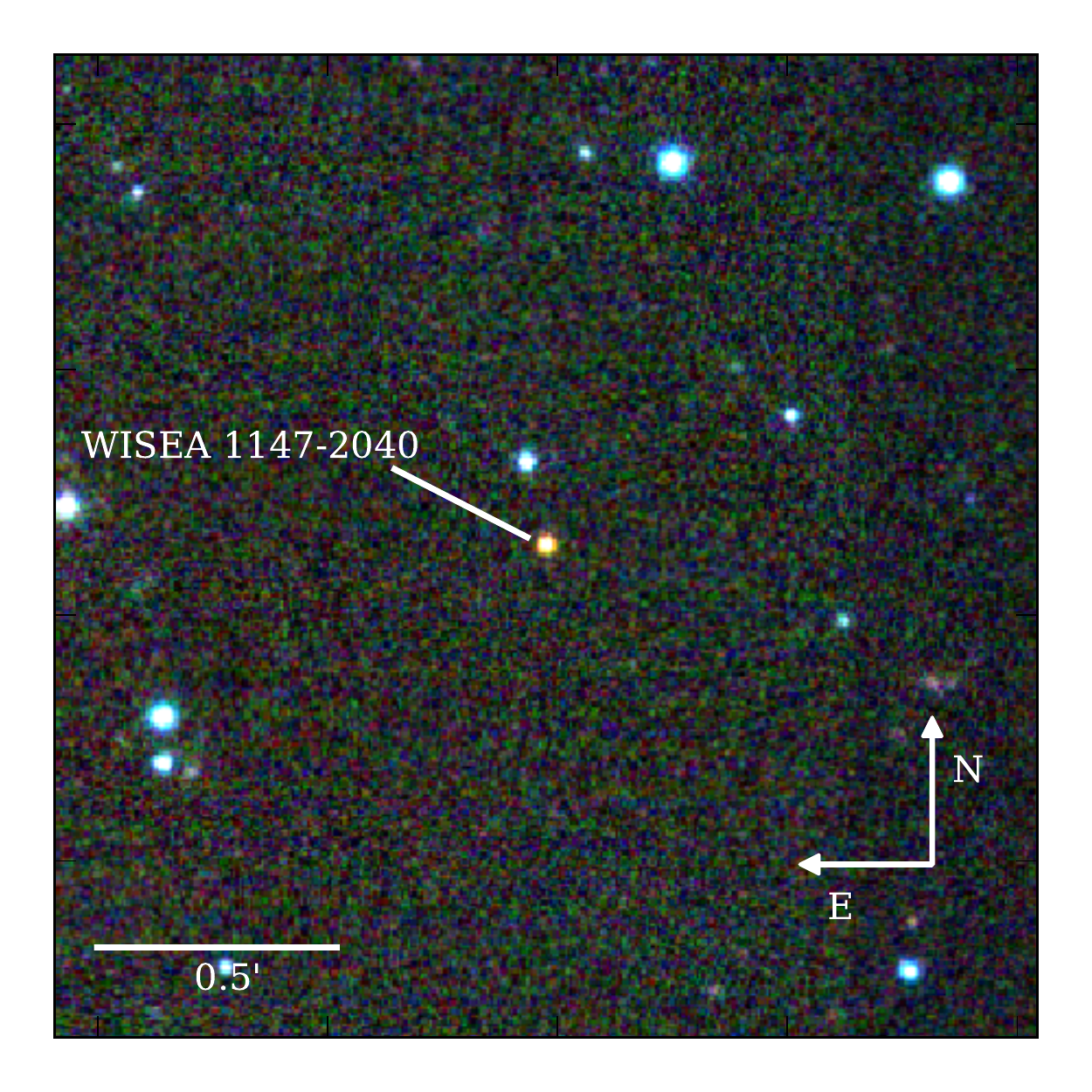}
\caption{Three color composite VHS image centered on the position of WISEA 1147$-$2040 ($Y$ = blue, $J$ = green, $K_{\rm S}$ = red).}  
\end{figure}

\section{Analysis}

\subsection{Spectral Typing}

We determined a spectral type for WISEA 1147$-$2040 following the method outlined in the Appendix of \cite{schneid14}.  A comparison of the SpeX spectrum of WISEA 1147$-$2040 with several near-infrared spectral standards \citep{kirk10} from the Spex Prism Spectral Library (SPL, \citealt{burg14}) is shown in the left panel of Figure 2.  While none of the L-type standards provide a good match to the spectrum of WISEA 1147$-$2040, the $J$-band portion most closely resembles that of the L7 standard.  We also compare the spectrum of WISEA 1147$-$2040 to the known, young, late-type L dwarfs 2MASS J03552337$+$1133437 (\citealt{reid06}, \citealt{fah13}),  WISEP J004701.06$+$680352.1 \citep{gizis12}, and WISE J174102.78$-$464225.5 \citep{schneid14} in the right panel of Figure 2.  WISEA 1147$-$2040 best matches the spectrum of WISE J174102.78$-$464225.5, an L7 type brown dwarf with likely membership in the $\beta$ Pictoris or AB Doradus moving groups \citep{schneid14}.  Note that \cite{gagne15} type WISE J174102.78$-$464225.5 as L5:$-$L7:$\gamma$.  Based on these comparisons, we estimate a spectral type of L7 $\pm$ 1 (very red) for WISEA 1147$-$2040.    

\begin{figure*}
\plotone{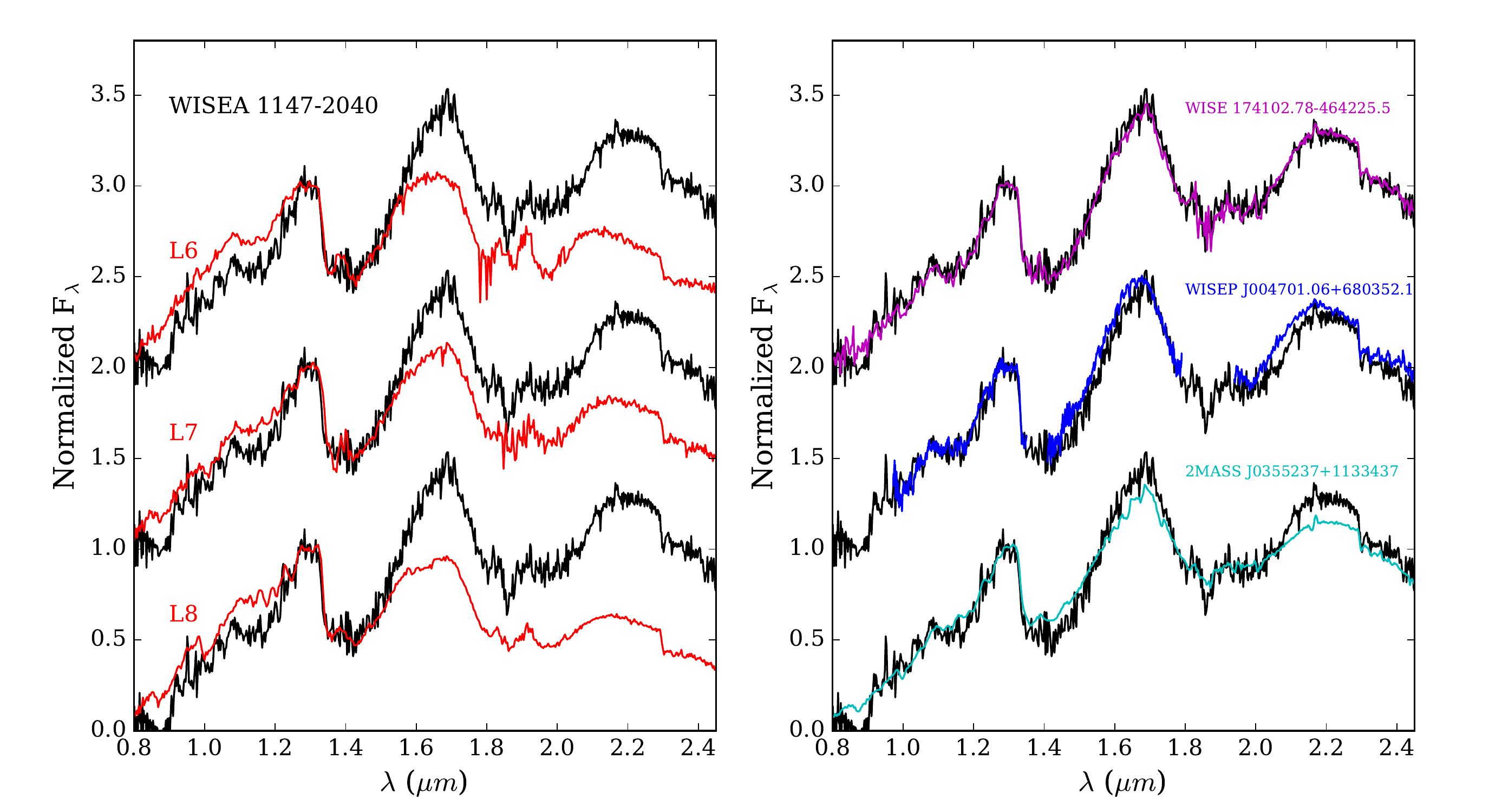}
\caption{{\it Left:} The IRTF/SpeX spectrum of WISEA 1147$-$2040 (black) compared with the near-infrared L6 (2MASSI J1010148$-$040649; \citealt{reid06}), L7 (2MASSI J0103320$+$193536; \citealt{cruz04}), and L8 (2MASSW J1632291$+$190441; \citealt{burg07}) standards (red). {\it Right:} The IRTF/SpeX spectrum of WISEA 1147$-$2040 (black) compared with 2MASS J03552337$+$1133437 (L5$\gamma$; \citealt{fah13}), WISEP J004701.06$+$680352.1 (L7 INT-G; \citealt{gizis15}), and WISE J174102.78$-$464225.5 (L7 (very red); \citealt{schneid14}). Each spectrum is normalized by the mean flux from 1.27 to 1.32 $\mu$m. }  
\end{figure*}

\subsection{Evidence of Youth}

WISEA 1147$-$2040 has several spectroscopic and photometric traits that provide strong evidence of its young age.  Because young brown dwarfs are still contracting to their final radii, they have lower surface gravities than field age brown dwarfs with the same mass.  One of the consequences thought to be due to low surface gravity is an unusually dusty atmosphere.  Such excessively dusty atmospheres then give rise to very red near-infrared colors.  WISEA 1147$-$2040 has an extremely red $J - K_{\rm S}$ color (2.57 $\pm$ 0.03 mag) determined from its VHS photometry.  In fact, only three free floating L dwarfs are known to have redder $J - K_{\rm S}$ colors; the 20 Myr old $\beta$ Pictoris moving group member PSO J318.5338$-$22.8603 ($J - K_{\rm S}$ = 2.837 mag; \citealt{liu13}), the extremely dusty L7 dwarf ULAS J222711$-$004547 ($J - K_{\rm S}$ = 3.04 mag; \citealt{mar14}), and the recently discovered TWA candidate member 2MASS J11193254$-$1137466 ($J - K_{\rm S}$ = 2.62 mag; \citealt{kell15}). While there are a few examples of L dwarfs with very red near-infrared colors that are not young (e.g., \citealt{kirk10}), bona fide young L dwarfs are typically found to have redder near-infrared colors than field dwarfs of the same spectral type (\citealt{cruz09}, \citealt{fah13}).  WISEA 1147$-$2040's red $J - K_{\rm S}$ color is the first indication that it is young.            

The near-infrared spectrum of WISEA 1147$-$2040 has a distinctly peaked $H$-band appearance and a steeper $K$-band slope.  For a field age brown dwarf with a normal surface gravity, the $H$ and $K$ band portions of their spectra are predominantly shaped by H$_2$O and collisionally-induced absorption (CIA) of $H_2$.  For a young brown dwarf, where the surface gravity is much lower, the effects of CIA are greatly reduced, resulting in the triangular $H$-band and steeper $K$-band shapes seen in their spectra \citep{rice11}.  \cite{allers13} defined the $H$-cont index to assess how peaked the $H$-band portion of a spectrum is.  We measure an $H$-cont index value of 0.968 for WISEA 1147$-$2040, which is decidedly different than the field L dwarf population, instead aligning well with other low gravity objects (see Figure 23 of \citealt{allers13} and Figure 5 of \citealt{gagne15}).  \cite{canty13} defined the H$_2$($K$) index as a measure of the $K$-band continuum shape and showed it could easily distinguish young brown dwarfs from field brown dwarfs for late-M spectral types.  \cite{schneid14} extended the H$_2$($K$) index into the L dwarf regime and showed that it could also be used to distinguish low gravity for L spectral types.  We measure an H$_2$($K$) value for WISEA 1147$-$2040 of 1.035, again aligning well with other low surface gravity L dwarfs (see Figure 10 of \citealt{schneid14} and Figure 14 of \citealt{gagne15}).  Other low gravity indices (e.g., FeH$_z$, VO$_z$) are only functional for spectral types $\lesssim$L5 and are therefore unsuitable for WISEA 1147$-$2040 \citep{allers13}.  

We can also find evidence of whether or not WISEA 1147$-$2040 has a low surface gravity by comparing its near-infrared spectrum to models with varying surface gravities.  We compare the atmospheric models of \cite{allard12} using the method of \cite{cush08}.  We find a best fitting temperature of 1500 $\pm$ 100 K and surface gravity of 4.0 $\pm$ 0.5, a surface gravity much lower than a typical field L dwarf.  The best-fitting model is shown in Figure 3.  The combination of an extremely red $J - K_{\rm S}$ color, triangular $H$-band shape, steeper $K$-band slope, and a low surface gravity estimate from model fitting lead us to conclude that WISEA 1147$-$2040 has a low surface gravity and is therefore young.  \cite{kirk08} show that low-gravity features only manifest in objects with ages less than that of the Pleiades, so we take $\sim$100 Myr as the upper age limit for WISEA 1147$-$2040.              






\begin{figure}
\plotone{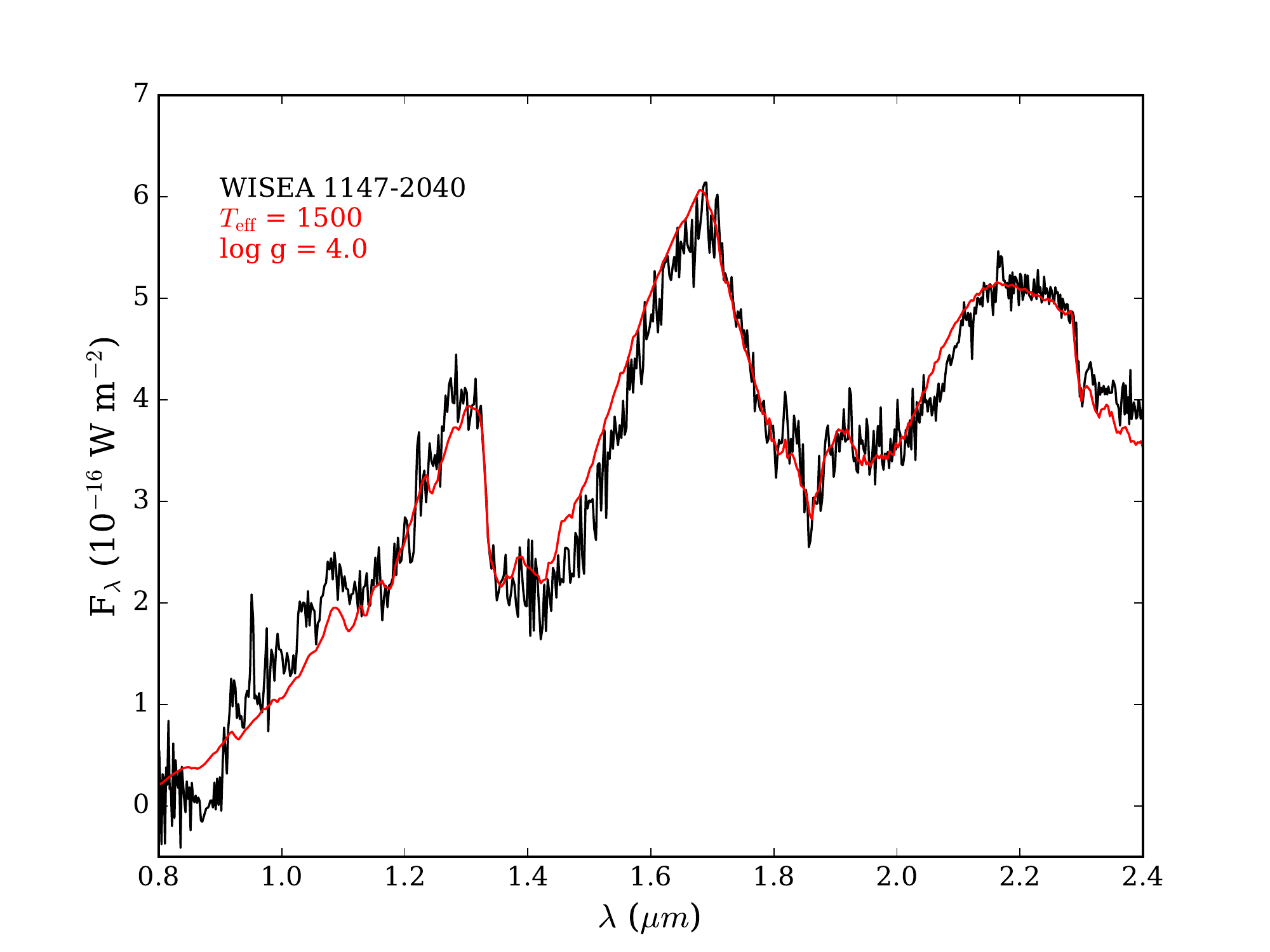}
\caption{The IRTF/SpeX spectrum of WISEA 1147$-$2040 (black) compared with the best fitting BT-Settl model (red; $T_{\rm eff}$ = 1500 K, log g = 4.0).}  
\end{figure}

\subsection{Membership in the TW Hya Association}

Because WISEA 1147$-$2040 shows youthful characteristics, we can now evaluate whether or not it belongs to one of the young nearby associations.  We first use the BANYAN II (\citealt{malo13}, \citealt{gagne14}) moving group membership evaluation tool, which takes the position and proper motion of a source and, through the use of a naive Bayesian classifier analysis, assesses membership probabilities for several nearby, young moving groups.  According to BANYAN II, WISEA 1147$-$2040 has an 84.32\% chance of belonging to TWA, under the assumption that it is young. 

We can first evaluate the feasibility of TWA membership for WISEA 1147$-$2040 by inspecting its sky position relative to other TWA members, as this particular association is confined to a particular area of the sky (compared to most other young moving groups).  The top panel in Figure 4 shows the position and proper motion vector of WISEA 1147$-$2040 along with confirmed TWA members from \cite{schneid12a}, \cite{schneid12b}, and \cite{murph15}, as well as high probability ($>$50\%) candidate TWA members from \cite{gagne15} and the recently announced L7 candidate TWA member 2MASS J11193254$-$1137466 from \cite{kell15}.  The figure shows that WISEA 1147$-$2040 is in close proximity to the other members of TWA, and is thus viable as a TWA candidate member.  BANYAN II also provides predicted radial velocity and distance values, assuming WISEA 1147$-$2040 is a TWA member, of 9.61 km s$^{-1}$ and 31.3 $\pm$ 3.8 pc, respectively.  While a higher resolution spectrum will be required to measure the radial velocity of WISEA 1147$-$2040, we can compare photometric and kinematic distance estimates to those predicated by BANYAN II to see if they are in agreement.  

We estimate the distance to WISEA 1147$-$2040 in two ways.  First, we estimate its distance photometrically.  Note that because WISEA 1147$-$2040 has such red near-infrared colors, the absolute magnitude-spectral type relations for field brown dwarfs cannot be used for WISEA 1147$-$2040 with all available photometric bands.  However, \cite{fah13} show that the young, very red L5$\gamma$ brown dwarf 2MASS J035523.37$+$113343.7 and the field L5 dwarf 2MASS J1507476$-$162738 have very similar absolute flux values at the $K$-band, while 2MASS J035523.37$+$113343.7 emits less flux than 2MASS J1507476$-$162738 at wavelengths shorter than $K$, and emits more flux at wavelengths redward of $K$.  We investigated whether or not this trait applies to other young, very red L7 dwarfs with measured parallaxes, specifically PSO J318.5338$-$22.8603 \citep{liu13} and WISEP J004701.06$+$680352.1 \citep{gizis12}.  Using the absolute magnitude-spectral type relations of \cite{dup12}, we calculate a $K_{\rm MKO}$ photometric distance for PSO J318.5338$-$22.8603 and a spectral type of L7 of 25.1 $\pm$ 0.2 pc, which agrees quite well with the measured parallax distance of 24.6 $\pm$ 1.4 pc \citep{liu13}, especially when compared to the distances found using the $J_{\rm MKO}$ (39.5 $\pm$ 0.7 pc) and W2 (17.9 $\pm$ 0.2 pc) band relations.  Similarly, for WISEP J004701.06$+$680352.1 and a spectral type of L7, we calculate a $K_{\rm S}$ photometric distance of 13.5 $\pm$ 0.2 pc, agreeing well with the measured distance of 12.2 pc \citep{gizis15}.  Again, the $K_{\rm S}$ photometric distance is much more accurate than the $J$ (18.5 $\pm$ 0.6 pc) and W2 (10.3 $\pm$ 0.1 pc) band photometric distances.  Therefore, we use the K$_{\rm S}$ absolute magnitude-spectral type relation of \cite{dup12} to estimate a photometric distance to WISEA 1147$-$2040 of 31.2 $\pm$ 1.5 pc.

We also estimate a kinematic distance to WISEA 1147$-$2040 following the ``moving cluster'' or ``convergent point'' method outlined in \cite{mam05}.  This method works because the proper motions of comoving  stars appear to converge to a single point on the celestial sphere from Earth's frame of reference.  \cite{mam05} provides coordinates of the convergent point for TWA of ($\alpha_{\rm cp}$ = 103.2 $\pm$ 1.5, $\delta_{\rm cp}$ = -30.7 $\pm$ 1.5), while \cite{duc14} find ($\alpha_{\rm cp}$ = 102.4 $\pm$ 1.4, $\delta_{\rm cp}$ = -27.3 $\pm$ 0.6).  For this analysis, we evaluate the kinematic distance to WISEA 1147$-$2040 using both sets of convergent point coordinates.  We use mean $UVW$ space velocities for TWA members of ($U, V, W$) = (-11.12 $\pm$ 0.90, -18.88 $\pm$ 1.56, -5.63 $\pm$ 2.78) km s$^{-1}$ from \cite{gagne14}.  For WISEA 1147$-$2040, we find kinematic distances of 32.6 pc and 32.0 pc for the \cite{mam05} and \cite{duc14} convergent points, respectively.  Radial velocities can also be estimated from the convergent point method, for which we find values of 9.2 and 8.7 km s$^{-1}$ for the \cite{mam05} and \cite{duc14} convergent points, respectively.  These radial velocity values agree very well with those estimated from BANYAN II.

Both the photometric distance from the $K_{\rm S}$ magnitude and the kinematic distance estimates of WISEA 1147$-$2040 agree remarkably well with each other, as well as the predicted distance from BANYAN II.  If we use either the photometric or kinematic distance to WISEA 1147$-$2040 as an additional input into BANYAN II, we find TWA membership probabilities of $\sim$96\%.  Using the BANYAN II predicted distance, we compare the galactic $XYZ$ coordinates of WISEA 1147$-$2040 with bona fide and high probability candidate members of TWA in the bottom panel of Figure 4.  Measured parallaxes for TWA members come from \cite{wein13} and \cite{duc14}, where distances with the smallest uncertainties were chosen for objects found in both studies.  We use the kinematic distances to TWA 3A, TWA 6, TWA 30A, and TWA 30B from \cite{duc14} and the kinematic distances to TWA 33 and TWA 34 from \cite{schneid12b}.   We use the quoted BANYAN distances to TWA 35 and TWA 36, as well as the high probability candidates from Table 4 of \cite{gagne15} from \cite{murph15} and \cite{gagne15}, respectively.  We also include the L7 TWA candidate member 2MASS J11193254$-$1137466 from \cite{kell15}, using their distance estimate of $\sim$25 pc.  The figure shows that WISEA 1147$-$2040 has $XYZ$ positions consistent with other TWA members.  Based on WISEA 1147$-$2040's young age, sky position, and the excellent agreement between its photometric distance estimate, kinematic distance estimate, and its predicted distance from BANYAN II, we conclude that WISEA 1147$-$2040 is a member of the TW Hya association.                

\begin{figure*}
\plotone{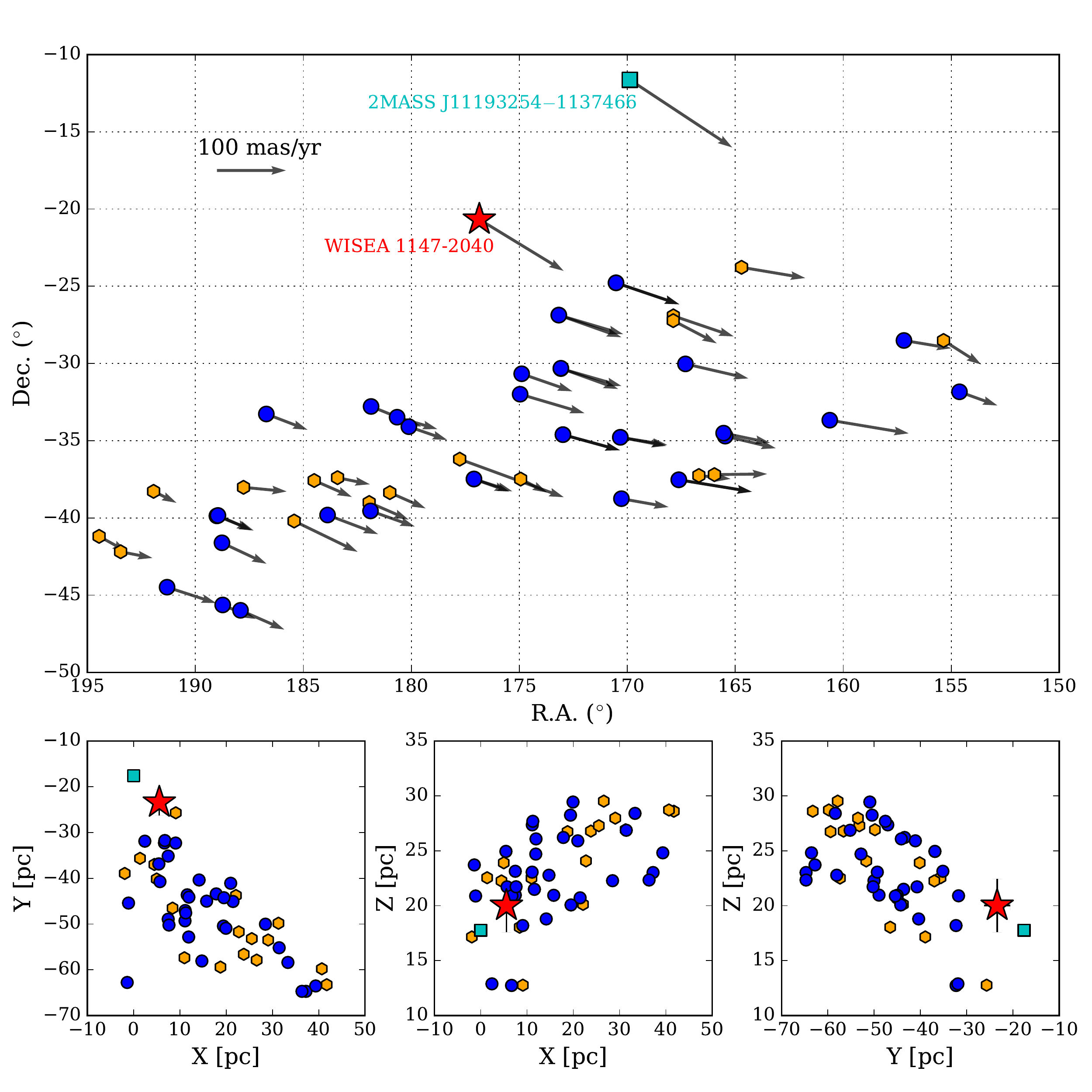}
\caption{{\it Top:} The position of WISEA 1147$-$2040 (red star) in the sky relative to known TWA members (blue circles), high probability ($>$50\%) candidates from \cite{gagne15} (orange hexagons), and 2MASS J11193254$-$1137466 (light blue square) from \cite{kell15}. {\it Bottom:} The XYZ positions of WISEA 1147$-$2040, known TWA members, high probability ($>$50\%) candidates from \cite{gagne15}, and 2MASS J11193254$-$1137466.  Symbols are the same as the upper panel. }  
\end{figure*}

\section{Discussion}

Based on WISEA 1147$-$2040's $T_{\rm eff}$ and log $g$ estimates from its spectrum, and an age of 10 $\pm$ 3 Myr for TWA \citep{bell15}, we estimate a mass of 6$-$13 $M_{\rm Jup}$ and 9$-$11 $M_{\rm Jup}$ from the COND evolutionary models \citep{bar03} and the \cite{sau08} ($f_{\rm sed}$ = 2) models, respectively.  Note that \cite{liu13} conclude that physical properties for PSO J318.5338$-$22.8603 (and other young, red L dwarfs) derived from near-infrared spectra are unreliable, instead using a combination of near-infrared spectra, photometry, and a distance measurement to determine a bolometric luminosity.  Mass, effective temperature, and surface gravity were then determined using the measured bolometric luminosity and moving group age.  

While a parallax measurement for WISEA 1147$-$2040 is yet unavailable, we can use the distance estimate from BANYAN II (31.3 $\pm$ 3.8 pc), combined with our flux calibrated near-infrared SpeX spectrum and {\it WISE} W1 and W2 magnitudes to estimate a preliminary bolometric luminosity.  Any wavelengths not covered by the SpeX near-infrared spectrum or AllWISE photometry we fill in with the best fitting model from Figure 2.  We Monte Carlo the uncertainties for the flux calibrated spectrum, AllWISE photometry, and distance estimate to determine the uncertainly of the calculated luminosity. We find log($L_{\rm bol}$/$L_{\odot}$) = -4.42 $\pm$ 0.11, which corresponds to $T_{\rm eff}$ values of $\sim$1100$-$1200 K at an age of 10 $\pm$ 3 Myr \citep{sau08}, much lower than the $T_{\rm eff}$ of WISEA 1147$-$2040 found via model fitting.  This luminosity corresponds to a mass estimate of 5$-$6 $M_{\rm Jup}$.  The differences between the parameters derived via fitting models to the near-infrared spectrum and those determined from the bolometric luminosity are similar to the differences seen for PSO J318.5338$-$22.860 in \cite{liu13}.    

Both the mass estimates from model fitting and from the bolometric luminosity make WISEA 1147$-$2040 the lowest mass free floating confirmed member of the TW Hya association and one of the lowest mass brown dwarfs in the Solar neighborhood.  In the TW Hya association, only the planetary mass companion 2M1207b (Chauvin et al.\ 2004, 2005) has a lower mass.  As such, WISEA 1147$-$2040 provides an exceptional laboratory for investigating the chemistry and cloud structure in young, planetary mass objects.  A higher resolution spectrum of WISEA 1147$-$2040 would provide both a radial velocity, helping to secure TWA membership, as well as additional gravity sensitive diagnostics (e.g., K I equivalent widths).  A trigonometric parallax for WISEA 1147$-$2040 would also further confirm it's membership in TWA. 


\acknowledgments

This publication makes use of data products from the {\it Wide-field Infrared Survey Explorer}, which is a joint project of the University of California, Los Angeles, and the Jet Propulsion Laboratory/California Institute of Technology, and NEOWISE, which is a project of the Jet Propulsion Laboratory/California Institute of Technology. WISE and NEOWISE are funded by the National Aeronautics and Space Administration.  This publication makes use of data products from the Two Micron All Sky Survey, which is a joint project of the University of Massachusetts and the Infrared Processing and Analysis Center/California Institute of Technology, funded by the National Aeronautics and Space Administration and the National Science Foundation.  This publication made use of observations obtained as part of the VISTA Hemisphere Survey, ESO Progam, 179.A-2010 (PI: McMahon).  


\begin{thebibliography}{}
\bibitem[Allard et al.(2012)]{allard12} Allard, F., Homeier, D., \& Freytag, B.\ 2012, Philosophical Transactions of the Royal Society of London Series A, 370, 2765 
\bibitem[Allers \& Liu(2013)]{allers13} Allers, K.~N., \& Liu, M.~C.\ 2013, \apj, 772, 79 
\bibitem[Baraffe et al.(2003)]{bar03} Baraffe, I., Chabrier, G., Barman, T.~S., Allard, F., \& Hauschildt, P.~H.\ 2003, \aap, 402, 701 
\bibitem[Bell et al.(2015)]{bell15} Bell, C.~P.~M., Mamajek, E.~E., \& Naylor, T.\ 2015, \mnras, 454, 593 
\bibitem[Bonnefoy et al.(2015)]{bonn15} Bonnefoy, M., Zurlo, A., Baudino, J.~L., et al.\ 2015, arXiv:1511.04082 
\bibitem[Bowler et al.(2014)]{bowl14} Bowler, B.~P., Liu, M.~C., Kraus, A.~L., \& Mann, A.~W.\ 2014, \apj, 784, 65 
\bibitem[Burgasser(2007)]{burg07} Burgasser, A. J. 2007b, \apj, 659, 655
\bibitem[Burgasser(2014)]{burg14} Burgasser, A.~J.\ 2014, Proceedings of the 2013 International
Workshop on Spectral Stellar Libraries, arXiv:1406.4887  
\bibitem[Canty et al.(2013)]{canty13} Canty, J.~I., Lucas, P.~W., Roche, P.~F., \& Pinfield, D.~J.\ 2013, \mnras, 435, 2650 
\bibitem[Chauvin et al.(2004)]{chau04} Chauvin, G., Lagrange, A.-M., Dumas, C., et al.\ 2004, \aap, 425, L29 
\bibitem[Chauvin et al.(2005)]{chau05} Chauvin, G., Lagrange, A.-M., Dumas, C., et al.\ 2005, \aap, 438, L25 
\bibitem[Cruz et al.(2004)]{cruz04} Cruz, K. L., Burgasser, A. J., Reid, I. N., \& Liebert, J. 2004, \apj, 604, L61
\bibitem[Cruz et al.(2009)]{cruz09} Cruz, K.~L., Kirkpatrick, J.~D., \& Burgasser, A.~J.\ 2009, \aj, 137, 3345 
\bibitem[Cushing et al.(2004)]{cush04} Cushing, M. C., Vacca, W. D., \& Rayner, J. T. 2004, \pasp, 116, 362
\bibitem[Cushing et al.(2008)]{cush08} Cushing, M.~C., Marley, M.~S., Saumon, D., et al.\ 2008, \apj, 678, 1372 
\bibitem[Ducourant et al.(2014)]{duc14} Ducourant, C., Teixeira, R., Galli, P.~A.~B., et al.\ 2014, \aap, 563, A121 
\bibitem[Dupuy \& Liu(2012)]{dup12} Dupuy, T.~J., \& Liu, M.~C.\ 2012, \apjs, 201, 19 
\bibitem[Emerson et al.(2004)]{emer04} Emerson, J.~P., Sutherland, W.~J., McPherson, A.~M., et al.\ 2004, The Messenger, 117, 27 
\bibitem[Faherty et al.(2013)]{fah13} Faherty, J.~K., Rice, E.~L., Cruz, K.~L., Mamajek, E.~E., \& N{\'u}{\~n}ez, A.\ 2013, \aj, 145, 2 
\bibitem[Gagn{\'e} et al.(2014a)]{gagne14} Gagn{\'e}, J., Lafreni{\`e}re, D., Doyon, R., Malo, L., \& Artigau, {\'E}.\ 2014, \apj, 783, 121 
\bibitem[Gagn{\'e} et al.(2015)]{gagne15} Gagn{\'e}, J., Faherty, J.~K., Cruz, K.~L., et al.\ 2015, \apjs, 219, 33 
\bibitem[Gauza et al.(2015)]{gauza15} Gauza, B., B{\'e}jar, V.~J.~S., P{\'e}rez-Garrido, A., et al.\ 2015, \apj, 804, 96 
\bibitem[Gizis et al.(2012)]{gizis12} Gizis, J.~E., Faherty, J.~K., Liu, M.~C., et al.\ 2012, \aj, 144, 94 
\bibitem[Gizis et al.(2015)]{gizis15} Gizis, J.~E., Allers, K.~N., Liu, M.~C., et al.\ 2015, \apj, 799, 203 
\bibitem[Hinkley et al.(2015)]{hink15} Hinkley, S., Bowler, B.~P., Vigan, A., et al.\ 2015, \apjl, 805, L10 
\bibitem[Kellogg et al.(2015)]{kell15} Kellogg, K., Metchev, S., Gei{\ss}ler, K., et al.\ 2015, \aj, 150, 182 
\bibitem[Kirkpatrick et al.(2008)]{kirk08} Kirkpatrick, J.~D., Cruz, K.~L., Barman, T.~S., et al.\ 2008, \apj, 689, 1295 
\bibitem[Kirkpatrick et al.(2010)]{kirk10} Kirkpatrick, J.~D., Looper, D.~L., Burgasser, A.~J., et al.\ 2010, \apjs, 190, 100 
\bibitem[Liu et al.(2013)]{liu13} Liu, M.~C., Magnier, E.~A., Deacon, N.~R., et al.\ 2013, \apjl, 777, L20 
\bibitem[Malo et al.(2013)]{malo13} Malo, L., Doyon, R., Lafreni{\`e}re, D., et al.\ 2013, \apj, 762, 88 
\bibitem[Mamajek(2005)]{mam05} Mamajek, E.~E.\ 2005, \apj, 634, 1385 
\bibitem[Marocco et al.(2014)]{mar14} Marocco, F., Day-Jones, A.~C., Lucas, P.~W., et al.\ 2014, \mnras, 439, 372 
\bibitem[Murphy et al.(2015)]{murph15} Murphy, S.~J., Lawson, W.~A., \& Bento, J.\ 2015, \mnras, 453, 2220 
\bibitem[Rayner et al.(2003)]{ray03} Rayner, J. T., Toomey, D. W., Onaka, P. M., et al. 2003, \pasp, 115, 362
\bibitem[Reid et al.(2006)]{reid06} Reid, I. N., Lewitus, E., Burgasser, A. J., \& Cruz, K. L. 2006, \apj, 639, 1114
\bibitem[Rice et al.(2011)]{rice11} Rice, E.~L., Faherty, J.~K., Cruz, K., et al.\ 2011, 16th Cambridge Workshop on Cool Stars, Stellar Systems, and the Sun, 448, 481 
\bibitem[Saumon \& Marley(2008)]{sau08} Saumon, D., \& Marley, M.~S.\ 2008, \apj, 689, 1327 
\bibitem[Schneider et al.(2012a)]{schneid12a} Schneider, A., Melis, C., \& Song, I.\ 2012, \apj, 754, 39 
\bibitem[Schneider et al.(2012b)]{schneid12b} Schneider, A., Song, I., Melis, C., Zuckerman, B., \& Bessell, M.\ 2012, \apj, 757, 163 
\bibitem[Schneider et al.(2014)]{schneid14} Schneider, A.~C., Cushing, M.~C., Kirkpatrick, J.~D., et al.\ 2014, \aj, 147, 34 
\bibitem[Skrutskie et al.(2006)]{skrut06} Skrutskie, M.~F., Cutri, R.~M., Stiening, R., et al.\ 2006, \aj, 131, 1163 
\bibitem[Vacca et al.(2003)]{vacca03} Vacca, W.~D., Cushing, M.~C., \& Rayner, J.~T.\ 2003, \pasp, 115, 389 
\bibitem[Weinberger et al.(2013)]{wein13} Weinberger, A.~J., Anglada-Escud{\'e}, G., \& Boss, A.~P.\ 2013, \apj, 762, 118 
\bibitem[Wright et al.(2010)]{wright10} Wright, E.~L., Eisenhardt, P.~R.~M., Mainzer, A.~K., et al.\ 2010, \aj, 140, 1868-1881 
\end{thebibliography}
\end{document}